\documentclass[article,twocolumn,preprintnumbers,superscriptaddress,nofootinbib]{revtex4-1}

\usepackage{amssymb}

\usepackage{enumerate}
\usepackage{amssymb}
\usepackage{amsmath}
\usepackage{tikz}
\usepackage[compat=1.1.0]{tikz-feynman}
\usepackage{feynmf}
\usepackage{makecell}
\usepackage{slashed}
\usepackage{natbib}
\usepackage{graphicx}

\usepackage[pdftex,bookmarks,linktocpage,pdfpagelabels,plainpages=false,hyperfigures,linkcolor=blue,citecolor=blue]{hyperref} 
\hypersetup{colorlinks=true}

\usepackage{mathrsfs,amssymb}  
\usepackage{cancel}
\usepackage[normalem]{ulem}

\newcommand\be{\begin{equation}}
\newcommand\ee{\end{equation}}
\newcommand{\comment}[1]{}
\newcommand\bea{\begin{eqnarray}}
\newcommand\eea{\end{eqnarray}}

\begin{document}

\begin{flushright}
MI-TH-2018
\end{flushright}
\bibliographystyle{apsrev4-1}

\title{Inverse Primakoff Scattering as a Probe of Solar Axions at Liquid Xenon Direct Detection Experiments}

\author{James B.~Dent} 
\email{jbdent@shsu.edu}
\affiliation{Department of Physics, Sam Houston State University, Huntsville, TX 77341, USA}

\author{Bhaskar Dutta}
\email{dutta@physics.tamu.edu}
\affiliation{Mitchell Institute for Fundamental Physics and Astronomy,
Department of Physics and Astronomy, Texas A\&M University, College Station, TX 77843, USA}

\author{Jayden L.~Newstead}
\email{jnewstead@unimelb.edu.au}
\affiliation{ARC Centre of Excellence for Dark Matter Particle Physics, School of Physics, The University of Melbourne, Victoria 3010, Australia}

\author{Adrian Thompson}
\email{thompson@physics.tamu.edu}
\affiliation{Mitchell Institute for Fundamental Physics and Astronomy,
   Department of Physics and Astronomy, Texas A\&M University, College Station, TX 77843, USA}

\begin{abstract}

We show that XENON1T and future liquid xenon (LXe) direct detection experiments are sensitive to axions through the standard $g_{a\gamma}aF\Tilde{F}$ operators due to inverse-Primakoff scattering. This previously neglected channel significantly improves the sensitivity to the axion-photon coupling, with a reach extending to $g_{a\gamma} \sim 10^{-10}$ GeV$^{-1}$ for axion masses up to a keV, thereby extending into the region of heavier QCD axion models. This result modifies the couplings required to explain the XENON1T excess in terms of solar axions, opening a large region of $g_{a\gamma}$ - $m_a$ parameter space which is not ruled out by the CAST helioscope experiment and reducing the tension with the astrophysical constraints.
We explore the sensitivity to solar axions for future generations of LXe detectors which can exceed future helioscope experiments, such as IAXO, for a large region of parameter space.
\end{abstract}

\maketitle

\par Dark matter direct detection experiments, initially designed to search for WIMP-like dark matter, have been adapted more broadly as detectors of Beyond Standard Model (BSM) physics. Notable among the wide class of BSM physics searches at direct detection facilities is the extraordinary sensitivity to possible axion or axion-like particles\footnote{We will use the generic `axion' (and symbol $a$) to encompass both cases, with the modifier QCD added when dealing with specific models developed to address the strong CP problem of QCD.} coupling to Standard Model particles (SM) \cite{Akerib:2017uem,Fu:2017lfc,Abe:2018owy,Armengaud:2018cuy,Aprile:2019xxb,Wang:2019wwo,Aralis:2019nfa}. By examining electronic recoils produced by a solar axion flux through the detector, these searches have probed a variety of $a$-SM couplings including axion-electron, axion-photon, and axion-nucleon interactions.

\par Recently, the XENON1T collaboration announced an observed excess of electron recoils in their low energy (1-30 keV) data, with a rise above the background-only model occurring below 7~keV~\cite{Aprile:2020tmw}. The solar axion flux is predicted to reside mostly in this energy range, making it a well-motivated hypothesis for the excess. The collaboration showed that a solar axion model can fit the data with a 3.5$\sigma$ significance, which is reduced to 2.1$\sigma$ if an unconstrained tritium background is introduced in the fitting. 

XENON1T placed constraints in a three-dimensional confidence limit volume in the parameter space of the axion-electron coupling, $g_{ae}$, along with the products $g_{ae}g_{a\gamma}$ and $g_{ae}g_{\rm an}^{\rm eff}$, where $g_{a\gamma}$ and $g_{\rm an}^{\rm eff}$ characterize the strength of axions coupling to photons and nucleons, respectively. These constraints were shown to be competitive with (or exceeding in some regions) the axion helioscope experiment CAST~\cite{Barth:2013sma} and the xenon-based dark matter direct detection experiments LUX~\cite{Akerib:2017uem} and PandaX-II~\cite{Fu:2017lfc}. The preferred region for the solar axion interpretation of the XENON1T result is in severe tension with astrophysical bounds, as discussed in~\cite{DiLuzio:2020jjp,Athron:2020maw} (see~\cite{DiLuzio:2020wdo} for a recent review that includes updated astrophysical bounds).

The analysis calculated the expected event rates produced by a solar axion flux consisting of three components arising from each of the couplings mentioned above. The XENON1T analysis considered detection through axioelectric scattering, the axion-analog of the photoelectric effect, off Xe electrons. This scattering process depends only on the $g_{ae}$ axion coupling, and not on either $g_{a\gamma}$ or $g_{\rm an}^{\rm eff}$.
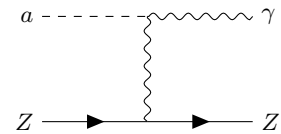
\begin{figure}[h]
 \centering
      \begin{tikzpicture}
              \begin{feynman}
         \vertex (o1);
         \vertex [right=1.4cm of o1] (f1) {\(\gamma\)};
         \vertex [left=1.4cm of o1] (i1){\(a\)} ;
         \vertex [below=1.4cm of o1] (o2);
         \vertex [right=1.4cm of o2] (f2) {\(Z\)};
         \vertex [left=1.4cm of o2] (i2) {\(Z\)};

         \diagram* {
           (i1) -- [scalar] (o1) -- [boson] (f1),
           (o1) -- [boson] (o2),
           (i2) -- [fermion] (o2),
           (o2) -- [fermion] (f2),
         };
        \end{feynman}
       \end{tikzpicture}
\caption{The inverse Primakoff process, where the axion $a$ coherently scatters with the electric fields of the entire atomic system $Z\equiv(e^-,N)$.}
    \label{fig:scattering}
\end{figure}

There is an alternative means of producing electron recoils through axion scattering that does not rely on the $g_{ae}$ coupling - namely through Primakoff scattering. In Primakoff scattering (also called the inverse Primakoff effect), shown in Fig.~\ref{fig:scattering}, an incident axion scatters off a charged particle through the $g_{a\gamma}$ coupling, producing an outgoing photon and recoil of the target particle. This channel occurs through a coherent interaction with the entire atomic form factor, not to be confused with a similar process involving the coherent interaction with external electromagnetic fields. The inverse Primakoff scattering process has been considered in several works~\cite{Avignone:1988bv,Creswick:1997pg,Creswick:2018stb,Avignone:1997th,Marsh:2014gca,Li:2015tsa}, including a recent analysis of the sensitivity of reactor neutrino experiments to axion-like particles with low-threshold detectors~\cite{Dent:2019ueq}.

In this work, we first describe the inverse Primakoff channel for axion detection in XENON1T (which can be applied to any direct detection experiments). We then explore the axion model parameter space for regions which can fit the XENON1T excess through Primakoff scattering within the detector. We demonstrate that the current XENON1T excess can be well-fit purely through a $g_{a\gamma}$ coupling (for both solar production and experimental detection). We show that there are regions of coupling and axion mass parameter space which fit the excess and are not ruled out by the CAST experiment. This region has constraints from HB stars which, however, can be evaded in the context of particle physics models~\cite{Jaeckel:2006xm,Khoury:2003aq,Masso:2005ym,Masso:2006gc,Dupays:2006dp, Mohapatra:2006pv,Brax:2007ak}. If this excess is instead due to an unmodelled background, we show the constraint emerging from this inverse Primakoff channel at the ongoing and future detectors.\\

{\bf{\emph{Models}-}} Peccei and Quinn (PQ) introduced a new global chiral symmetry into the Standard Model in order to solve the strong CP problem \cite{Peccei:1977hh}. This symmetry is broken dynamically and the resulting pseudo Nambu-Goldstone boson is the axion \cite{Weinberg:1977ma,Wilczek:1977pj}. Although axion couplings to all Standard Model particles can be considered, in the present work we examine the couplings to photons and leptons within the interaction Lagrangian
\begin{equation}
    \mathcal{L} \supset -\dfrac{1}{4}g_{a\gamma}aF_{\mu\nu}\Tilde{F}^{\mu\nu}+ig_{ae}a\bar{\psi}\gamma^5\psi
\end{equation}
These couplings are model dependent, with some models allowing lepton couplings only beginning at one-loop order. Within a given model, these couplings are not strictly independent, as loop effects can correlate them. However, in the present work we allow these parameters to be independently fit to the data. We will also comment on the common QCD axion models of Dine-Fischler-Srednicki-Zhitnitskii (DFSZ)~\cite{Dine:1981rt,Zhitnitsky:1980tq}, and Kim-Shifman-Vainshtein-Zakharaov (KSVZ)~\cite{Kim:1979if,Shifman:1979if}, which provide specific, model dependent forms for $g_{a\gamma}$ and $g_{ae}$. In addition, effective couplings to nucleons (of a similar form as the electron interaction term~\cite{Alessandria:2012mt}) will be included in the next section.

The explicit coupling correlations can be made to connect our phenomenological axion model to specific DFSZ and KSVZ QCD axion models in the analysis. For these we make use of the coupling relationships which can be found in refs.~\cite{Chang:1993gm,diCortona:2015ldu,Irastorza:2018dyq,Giannotti:2017hny}.
The KSVZ and DFSZ type axion models can be classified by the ratio anomaly parameters $E$ and $N$ specific to each variety of model. In addition, the definitions of $g_{a\gamma}$ and $g_{ae}$ are correlated in each model ($g_{ae}$ being loop-induced in KSVZ models, while $g_{a\gamma}$ is loop-induced in DFSZ), and depend on the QCD scale factor $\Lambda \sim 1$ GeV and mixing parameter $\tan\beta$ (just for DFSZ models). We assume $\tan\beta = 140$ (DFSZ I) and $\tan\beta = 0.28$ (DFSZ II), taken from fits to accommodate unitarity and stellar cooling~\cite{Giannotti:2017hny}. Although our analysis will permit $g_{a\gamma}$ and $g_{ae}$ to be free, we can use their correlations in the varieties of KSVZ and DFSZ models to pick out subspaces of the total parameter space. Various instances of these models have been explored over a range of $E/N$ values~\cite{DiLuzio:2017pfr}.\\

{\bf{\emph{Analysis}-}}\label{sec:analysis} We consider two possibilities of solar axion production (for details regarding solar axion fluxes, see Refs.~\cite{Redondo:2013wwa, Andriamonje:2007ew,Hagmann:2008zz, Moriyama:1995bz,Andriamonje:2009dx,Alessandria:2012mt, Haxton:1991pu,Alessandria:2012mt} and the Appendix) 
and subsequent scattering within the XENON1T volume. First, we examine Primakoff production in the sun, followed by photon production through Primakoff scattering in the detector (this combination is purely dependent on $g_{a\gamma}$).We note here that solely using this coupling approximates the phenomenology of $g_{a\gamma}$ dominated axion models, e.g., KSVZ-type axion models, for which axion-electron interactions happen through loop-induced processes, and would be suppressed. This also allows a direct comparison with other experiments, such as haloscopes and helioscopes, which have sensitivity to $g_{a\gamma}$ as a function of the axion mass, $m_a$. Second, we analyze a non-zero $g_{ae}$ in conjunction with a non-zero $g_{a\gamma}$, which allows for both ABC and Primakoff solar production, along with axioelectric and Primakoff scattering detection. This analysis is complementary to that of XENON1T which allowed for the possibility of Primakoff production, but only detection through the axion-electron scattering controlled by the $g_{ae}$ coupling. 

Inverse Primakoff scattering allows solar axions to coherently scatter from the Xe atomic electric field and back-convert into photons in the detector volume; $a Z \to \gamma Z$, proceeding through a $t$-channel photon exchange (Fig.~\ref{fig:scattering}). Therefore the inverse Primakoff scattering contributes to the axion hypothesis and must be included. The final state photon will have a short mean free path ($\sim \mu$m) and cause an electron-like recoil in the LXe TPC. This is verified by XENON1T's $^{37}$Ar calibration; $^{37}$Ar decays via electron capture of a K-shell electron to ${37}$Cl, the core-hole is filled from an outer shell electron, which emits a 2.8 keV photon with 90$\%$ branching fraction~\cite{Aprile:2020tmw,talk}. Additionally, the xenon response to low energy photons was measured by the LUX collaboration who found that, in the region of interest, the response is approximately the same as for electron recoils~\cite{Akerib_2017}. Therefore it is reasonable to assume that all inverse Primakoff events are detected with the same efficiency and energy resolution as electron recoils.

For an axion of momentum $k_a$, the inverse Primakoff cross section is given by~\cite{Avignone:1988bv,Creswick:1997pg,Avignone:1997th}
\begin{equation}
    \sigma(k_a) = \dfrac{Z^2 \alpha g_{a\gamma}^2}{2} \bigg(\dfrac{2 r_0^2 k_a^2 + 1}{4 r_0^2 k_a^2} \ln \Big(1+4r_0^2 k_a^2 \Big) - 1 \bigg)
    \label{eq:xs}
\end{equation}
where $r_0$ is the screening length for which we take as the Wigner-Seitz radius in LXe, $0.685$~\AA, given at the nominal XENON1T density\footnote{We have changed this value of the screening radius from 2.45~\AA ~\cite{Walters_2006}\, to 0.685~\AA \, in accordance with the more common expression for the Wigner-Seitz radius~\cite{Ashcroft}, which accounts for at most a factor of $10$ reduction in the cross section at 1 keV down to a factor of 3 reduction at 5 keV with respect to the old value.}. The parameter $r_0$ in Eq.~\ref{eq:xs} comes from an atomic form factor that assumes a screened Coulomb potential. However, another study on the inverse Primakoff scattering~\cite{Abe:2020vff} has illustrated that the choice of atomic form factor has a significant impact on the size of the cross section below $E_a = 10$ keV; another model using a relativistic Hartree-Fock treatment of the form factor gives a smaller cross section by a factor of $\mathcal{O}(10)$. While the screened Coulomb form factor is less sophisticated, the RHF method is based on isolated atomic potentials and does not account for varying screening lengths in liquid phase; since working out a complete model of the atomic response in liquid is beyond the scope of this work, we will present results for both form factors. We shall refer to the screened Coulomb form factor with the Wigner-Seitz screening radius as ``WS" and the Hartree-Fock form factor as ``RHF". However, we find that $g_{a\gamma}$ differs by at most a factor 2 due to these various choices of these form factors.

We consider inverse Primakoff scattering in addition to the axioelectric absorption process outlined in the analysis performed by XENON1T (see also~\cite{Dimopoulos:1985tm,Ljubicic:2004gt,PhysRevD.35.2752,Alessandria:2012mt}). Also, it is possible that axions undergo inverse Compton scattering off electrons at rest in LXe, $a e^- \rightarrow \gamma e^-$~\cite{Avignone:1988bv}, but this is a subdominant process ($\propto Z$) in comparison to axioelectric scattering ($\propto Z^5$). If both axion-photon and axion-electron couplings are present, there are interference terms present in the total matrix element of the combined processes, which are also subdominant, but we include them as a matter of completeness.

\begin{figure}
    \centering
    \includegraphics[width=0.45\textwidth]{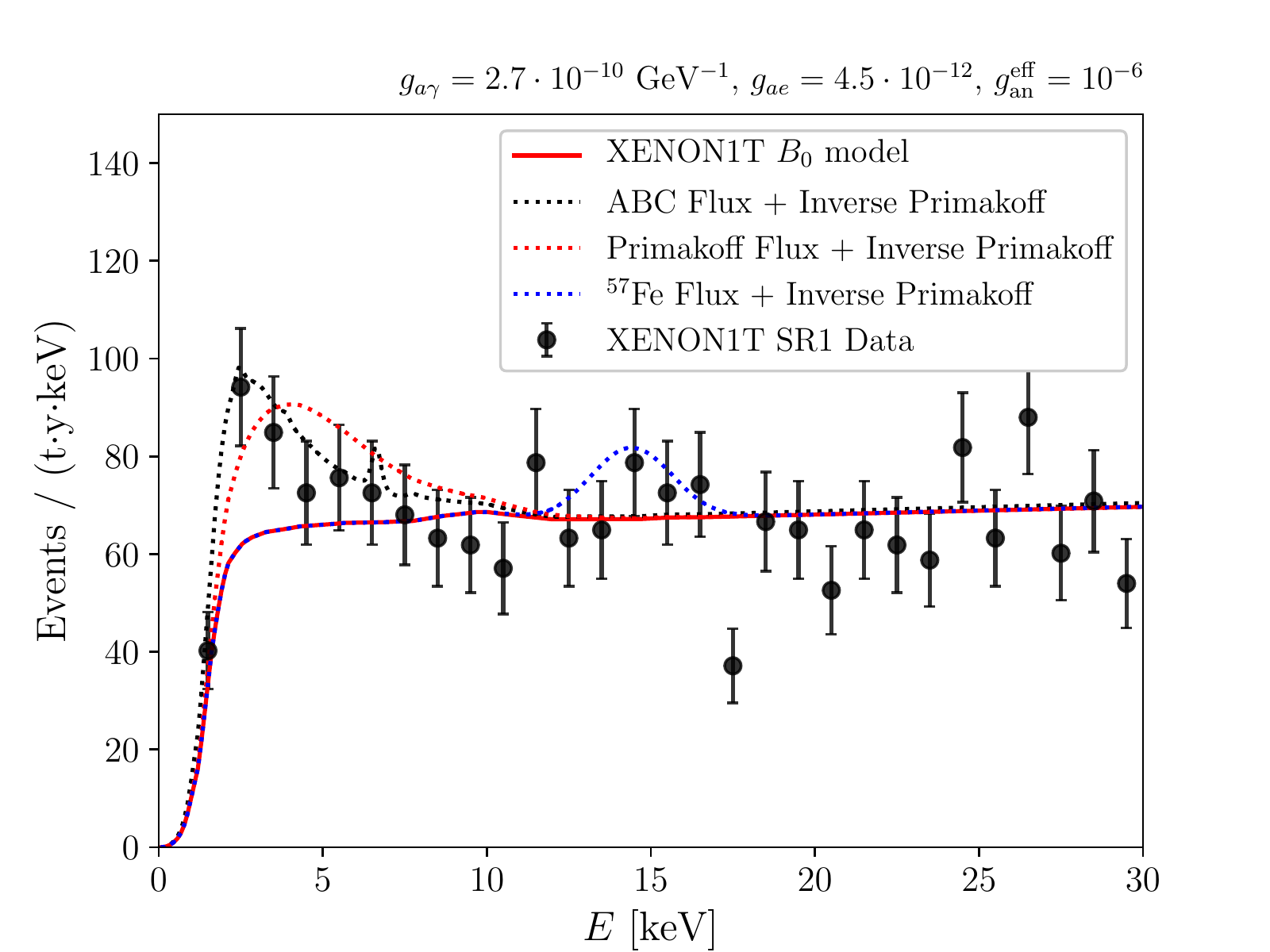}
    \caption{The event rate distributions for inverse Primakoff scattering in LXe for a ton-year exposure from Primakoff-produced, ABC-produced, and $^{57}$Fe-produced axions are shown for select choices of axion couplings, and added to the ``$B_0$" background model.}
    \label{fig:primakoff_spectra}
\end{figure}

To predict the event spectra from axions produced through ABC, Primakoff, and $^{57}$Fe, we convolve the fluxes in each case with the total cross sections, for inverse Primakoff scattering or axioelectric absorption, and multiply by the detector efficiency~\cite{Aprile:2020tmw}. In addition, we approximate the detector response for the energy resolution by convolving the simulated differential event distribution with an energy-dependent Gaussian smearing function~\cite{Aprile:2020tmw,Aprile:2020yad,Aprile:2019dme}. The event distribution for Primakoff-produced axions that undergo inverse Primakoff scattering in the LXe fiducial volume over a ton-year exposure is shown in Figure~\ref{fig:primakoff_spectra}.

We perform a likelihood analysis given the data and ``$B_0$" background hypothesis taken in~\cite{Aprile:2020tmw} and our axion signal hypotheses using the Bayesian inference package \texttt{MultiNest}~\cite{Feroz:2008xx,Feroz:2007kg,Feroz:2013hea}. A binned log-gaussian likelihood is constructed over bins $i$, with signal event rates $\{\mu_i \}$ and observed events $\{n_i\}$ ranging from 1 to 29 keV, taken with errors $\sigma_i$ reported by XENON1T. 

We wish to investigate several scenarios of signal and background models in the context of the excess, enumerated as follows: (I) Primakoff-produced axions detected through solely inverse Primakoff scattering, (II) Primakoff-produced and scattered axions with an additional $^3$H component, (III) the $^3$H component alone, repeating the methods used in the XENON1T analysis, (IV) all production mechanisms (ABC, Primakoff, $^{57}$Fe) and all scattering channels (Primakoff, axioelectric, Compton) allowed in the detector, and finally, (V) all flux components and scattering channels along with an unconstrained $^3$H component. Fits (I) and (II) aim to test the robustness of the Primakoff-only (pure-$g_{a\gamma}$) fit after introducing a $^3$H background component, while (III) validates the $^3$H-only fit. Fits (IV) and (V) aim to test the same robustness when all axion production and detection mechanisms are allowed. For each of these cases we will assume flat priors over appropriately large intervals on the free parameters ($g_{ae}$, $g_{a\gamma}$, and $g_{\rm an}^{\rm eff}$) in the likelihood scan. We keep the axion mass fixed below 100 eV, since the production and scattering rates remain unchanged in this limit, and keep $g_{a\gamma}$ and $g_{ae}$ sufficiently small as to avoid $a\to\gamma\gamma$ decays that would be ruled out several constraints.

We will also consider alternative scenarios where the low energy excess either disappears with more exposure at third-generation xenon experiments, or that the background model becomes more well-understood and shows no excess. We can simulate these possibilities to forecast future exclusions in parameter space. Future limits and the five cases that we consider for the analysis of the excess are discussed in the next section.\\

{\bf{\emph{Fit Results}-}}\label{sec:results}
After checking all five cases described in the previous section with the likelihood-ratio test statistic, we find that the $^3$H unconstrained model rejects the background-only hypothesis at a $2.3\sigma$ level, in agreement with the XENON1T result. When Primakoff production and detection mechanisms are added to the signal model that includes the unconstrained $^3$H component, we find a significance of $2.6\sigma$, while if we remove the $^3$H component and just include Primakoff production and detection, we reject the background-only hypothesis at $3.1\sigma$, slightly less significant than the XENON1T result which omitted the inverse Primakoff detection component. This may be intuitively understood by the shape difference between the Primakoff flux with inverse Primakoff response, shown as the red dotted curve in Fig.~\ref{fig:primakoff_spectra}, and the response from the ABC-produced axioelectric absorption which is peaked at lower energies more than the inverse Primakoff response. Finally, if we allow for all fluxes and detection channels that we considered to be present in the likelihood scan, we find a rejection of the background at a level of $3.7\sigma$, mildly higher than the XENON1T result, while if we also include an unconstrained $^3$H, the significance is reduced to $2.95\sigma$.

\begin{figure}
    \centering
    \includegraphics[width=0.5\textwidth]{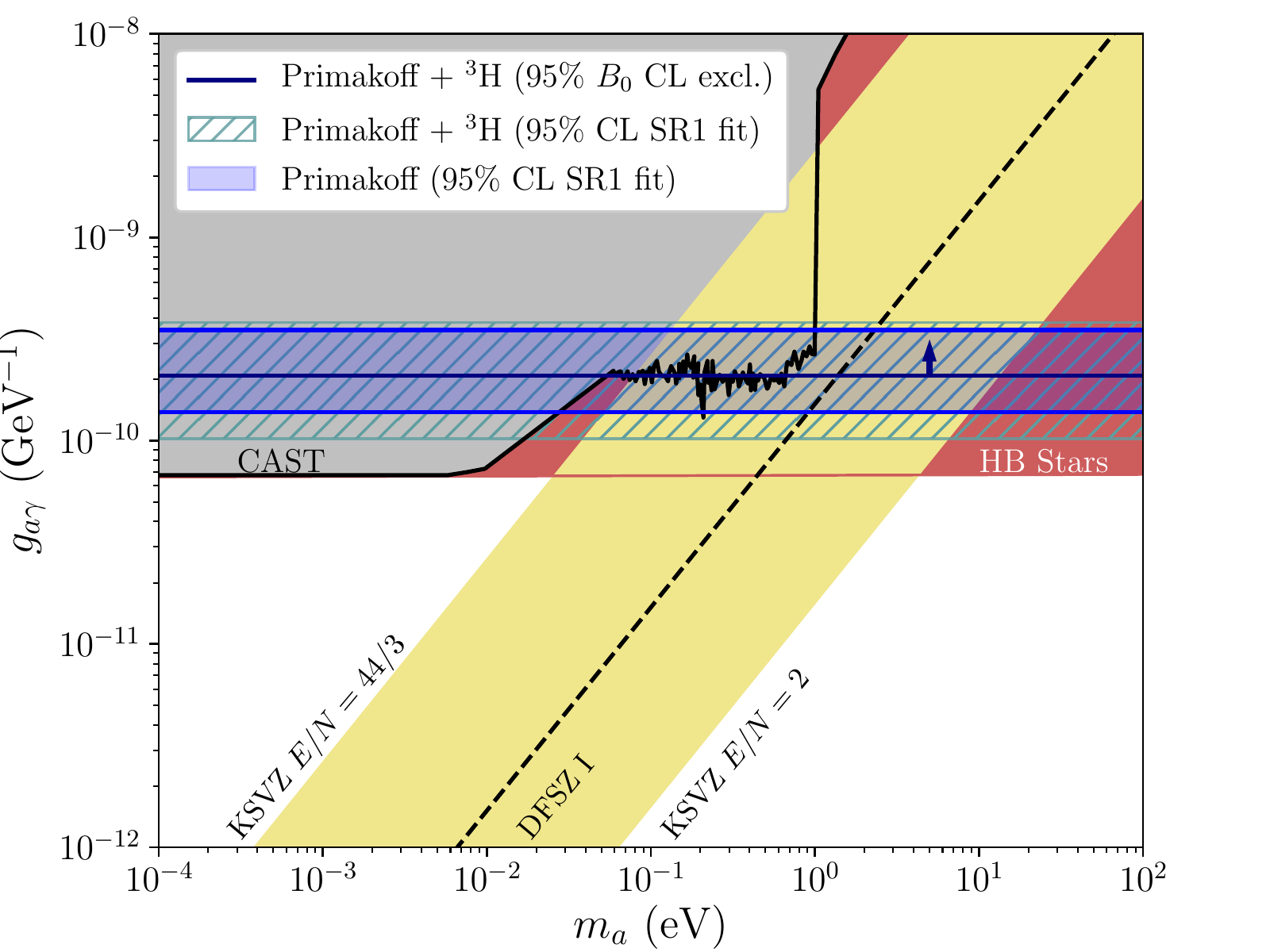}
    \caption{The fit with Primakoff-produced solar axions undergoing solely inverse-Primakoff scattering in XENON1T (blue band) is compared against the limit when $^3$H is included as a background on top of the signal hypothesis (hatched band) at 95\% CL. We also show the bound from the Primakoff signal hypothesis tested against the $B_0$ background, simulating a no-excess scenario. For these fits we take $r_0 = 0.685$~\AA,  corresponding to the Wigner-Seitz screening model. We discuss the existing constraints in the text.}
    \label{fig:fit_prim}
\end{figure}

For the purely Primakoff-driven production and detection scenario, in Fig.~\ref{fig:fit_prim} we display our best fit region in the $g_{a\gamma}-m_a$ parameter space for the XENON1T excess, as well as the current limits from the CAST helioscope and astrophysical bounds. The CAST limits~\cite{Anastassopoulos:2017ftl} provide a bound of ${g_{a\gamma}<0.66\times~10^{-10}\,{\rm GeV}^{-1}}$(95\% CL) for ${m_a<0.02\,{\rm{eV}}}$, and ${g_{a\gamma}<2\times~10^{-10}\,{\rm GeV}^{-1}}$ (95\% CL) for ${m_a<0.7\,{\rm{eV}}}$. The excess explanation evades the CAST constraint for $m_a>0.03$ eV. Not pictured are the other constraints of $g_{a\gamma} < 4.1\cdot10^{-10}$ GeV$^{-1}$, from a combined global analysis of helioseismology and solar neutrino arguments~\cite{Vinyoles:2015aba}, and $g_{a\gamma} \lesssim 6\cdot 10^{-10}$ GeV$^{-1}$ for $0.8 \lesssim m_a \lesssim 1.0$ eV, from SUMICO~\cite{Inoue:2008zp}, both of which are evaded by the XENON1T fit for all masses considered.

Bounds from the $R$-parameter - the ratio between the number of horizontal branch (HB) stars and red giant branch (RGB) stars in older stellar clusters~\cite{Ayala:2014pea,Giannotti_2016} - also sets a very stringent bound of ${g_{a\gamma}<0.6\times~10^{-10}{\rm GeV}^{-1}}$ (95\%~CL) for $g_{ae}$=0 (for ${g_{a\gamma}\sim 10^{-11}{\rm GeV}^{-1}}$, the 95\% CL region extends to $g_{ae}\sim2.6\times 10^{-13}$, as seen for example, in the analysis of \cite{Giannotti:2015kwo}) but extends to higher axion masses than the CAST bound. However, since HB and RGB stars have much higher density (by two to four orders of magnitude) and higher core temperatures (by a factor of seven) compared to the sun, mechanisms exist in the context of specific particle physics models which could allow the evasion of the bounds emerging from the null observation of axions associated with these astrophysical objects, e.g.~\cite{Jaeckel:2006xm,Khoury:2003aq,Masso:2005ym,Masso:2006gc,Dupays:2006dp, Mohapatra:2006pv,Brax:2007ak,gao2020reexamining,Bloch:2020uzh,DeRocco:2020xdt}.

\begin{figure}
    \centering
    \includegraphics[width=0.5\textwidth]{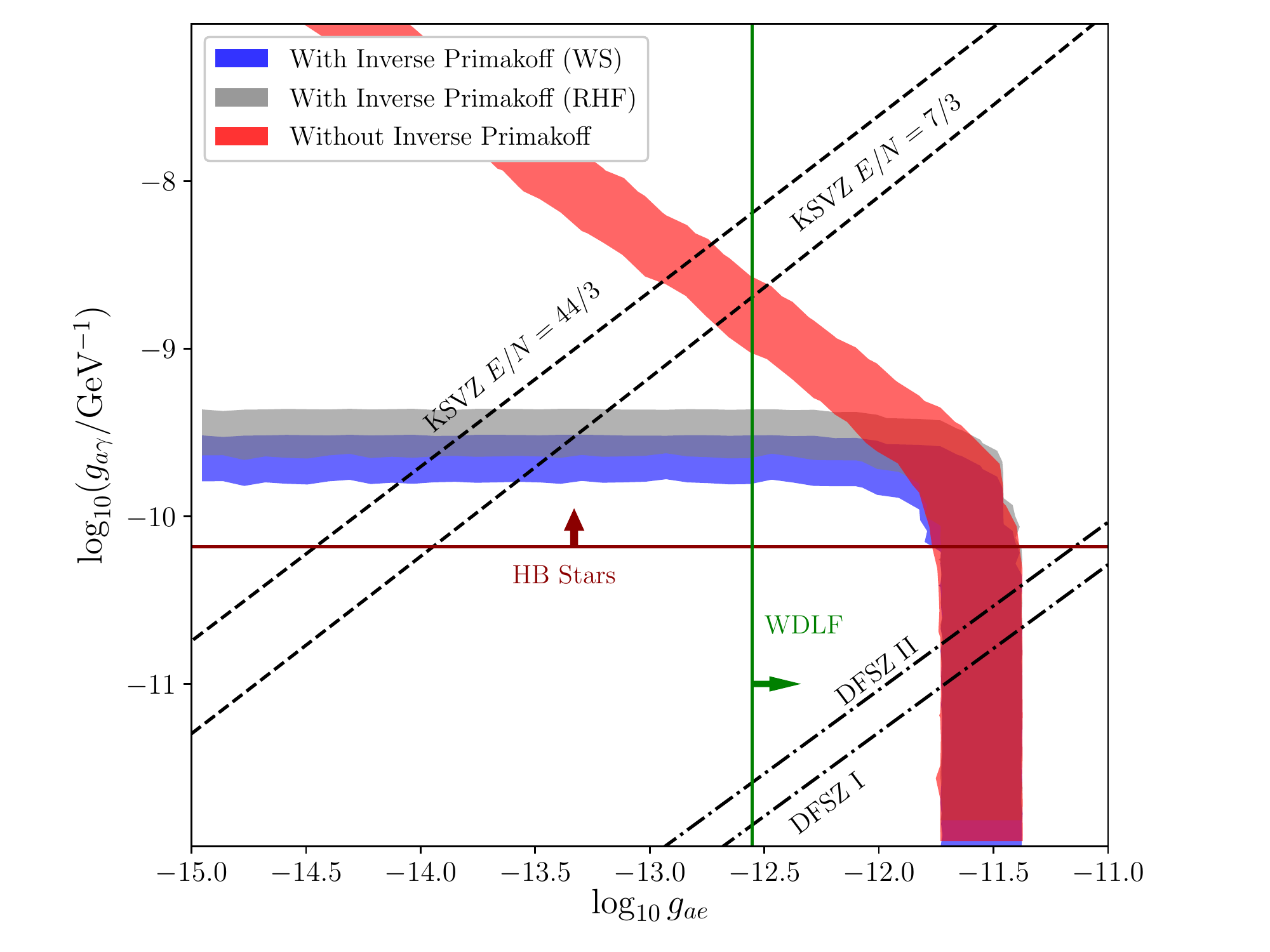}
    \includegraphics[width=0.5\textwidth]{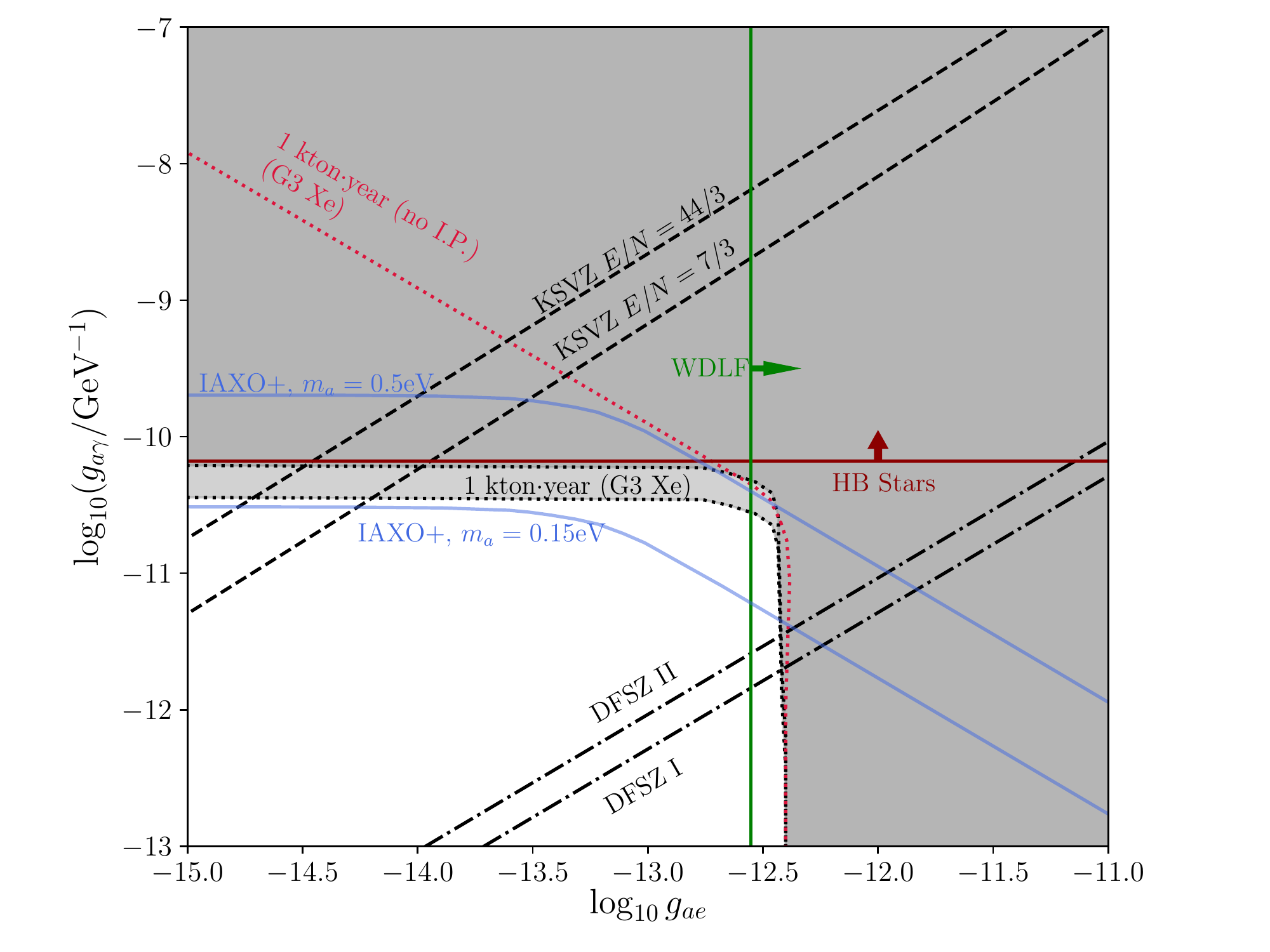}
    \caption{\textit{Top}: 95\% CL contours are shown for fits to the XENON1T excess for all axion flux components with only axioelectric scattering (red) and with both inverse-Primakoff and axioelectric scattering using the WS-screened form factor (blue) and the RHF form factor (gray). Here we consider $m_a= 0.7$ eV, however, the plot does not change for any $m_a<100$ eV and the CAST constraints are evaded for $m_a>0.03$ eV. \textit{Bottom}: Projections of the 95\% CL future exclusions (using the Wigner-Seitz FF in light gray and RHF form factor in dark gray) set by G3 Xe over a 1 kton$\cdot$year exposure given background-only observations. The exclusion line for 1 kton$\cdot$year without inverse Primakoff (I.P.) scattering is shown for comparison (dotted red). We also show the IAXO+ projection (blue) which begins to lose sensitivity for $m_a \gtrsim 0.01$ eV.
    }\label{fig:gaevsgagamma}
\end{figure}

The evasion could involve additional scalar degrees of freedom around the HB star temperature by invoking a phase transition~\cite{Mohapatra:2006pv}, or the axion as a chameleon-type field with its mass depending on the environmental matter density $\rho$~\cite{Khoury:2003aq,Brax:2007ak}. In addition, the possibility that the axion is a composite particle with a form factor has been explored~\cite{Masso:2005ym,Masso:2006gc,Dupays:2006dp}, leading to a suppression of the production in the HB stars, as well as models with a paraphoton where the axion-like particles are trapped in the HB star interior thus evading the stellar bounds. 

Another possibility considers a population of axions gravitationally bound to the Sun. In~\cite{VanTilburg:2020jvl}, it is shown that stellar emission of non-relativistic axions into gravitationally bound orbits can significantly increase the flux of axions on Earth. This additional flux reduces the coupling required to explain the XENON1T excess ($g_{ae}\sim10^{-13}$) and thus reduces tension with the astrophysical constraints. Further work is required to determine if this scenario can indeed provide a robust explanation of the XENON1T excess.

In Fig.~\ref{fig:gaevsgagamma} (top), we plot $g_{a\gamma}$ vs.~$g_{ae}$ where contributions from both axion-electron and axion-photon couplings are included. The red shaded regions show the XENON1T excess fit without considering inverse Primakoff while the blue shaded region utilizes inverse Primakoff. We find that the improvement in $g_{a\gamma}$ due to inverse Primakoff is quite significant for $g_{ae}\lesssim10^{-12}$, and one can see that the transition from the $g_{ae}$-dominated signal to the $g_{a\gamma}$-dominated signal occurs around $g_{ae}=10^{-12}$ and $g_{a\gamma}=10^{-10}$GeV$^{-1}$. In the limit of small $g_{ae}$ the inverse Primakoff channel provides flat sensitivity that is especially improved for KSVZ-type models. Constraints from white dwarf luminosity function (WDLF) place bounds on $g_{ae}<2.8\cdot10^{-13}$~\cite{Bertolami:2014wua}.

If the excess is due to a background phenomenon, the current data constrain the axion parameter space. We compute this constraint by testing our signal hypothesis against the $B_0$ model at various exposures; in Fig.~\ref{fig:fit_prim}, we show the constraint in $g_{a\gamma}$ as a function of $m_a$ and we find that the constraint is already better than the CAST constraint for $m_a>0.04$ eV. In Fig.~\ref{fig:gaevsgagamma} (bottom), we show the next-generation xenon (G3 Xe) constraint (with a 1 kton-year exposure~\cite{Szydagis:2016few}) and find that the 95\% CL can overcome even the HB stars constraint and start exploring the mild hint (2.4$\sigma$) region of stellar cooling within 1$\sigma$. Interestingly, this is only possible with the inclusion of the inverse Primakoff channel since without this channel the constraint could be worse by a few orders of magnitude. We also find that our projected sensitivity for a 1 kton$\cdot$year exposure at a G3 LXe experiment is competitive with future helioscope experiments. The proposed DARWIN detector would achieve a 200 ton-year exposure~\cite{Aalbers:2016jon}, thereby covering the current HB Stars constraint. We compare the 1 kton$\cdot$year projection against the projected sensitivities for IAXO+ with masses $m_a > 0.1$ eV, where sensitivity begins to diminish for larger masses~\cite{Armengaud:2019uso}. Additionally, future direct detection experiments with directional sensitivity would be able to use the directional information to reduce backgrounds and further increase their sensitivity to solar axions. This is especially useful in the Primakoff channel, where the axion's incoming direction is approximately preserved by the photon in the relativistic limit.\\

{\bf{\emph{Conclusion}- }}In this work, we investigated inverse Primakoff scattering as a new detection channel at liquid xenon based direct detection experiments. We showed that sole use of the coupling $g_{a\gamma}$ can fit the recent XENON1T excess. The fitting of the excess is free of the leading helioscope CAST constraint for $m_a\gtrsim 0.03$ eV. If this excess is due to the background we find that the 95\%CL exclusion limit is also better than the CAST limit. The tension associated with the astrophysical constraints (e.g., WD constraint) ruling out the axion interpretation combining $g_{ae}$ and $g_{a\gamma}$ gets relaxed with the inclusion of the inverse Primakoff effect, however, the discrepancy with the  $R$ parameter is still 8$\sigma$~\cite{DiLuzio:2020jjp}. However, these limits can be dependent on the given particle physics model. Additionally, next-generation xenon experiments can overcome the HB stars limit, and for $g_{ae}=10^{-13}$, the 2.4$\sigma$ hint region of stellar cooling can be probed within $1\sigma$. In addition, these future bounds would be applicable for masses $m_a < 1$ keV, covering complementary regions of parameter space for which future helioscopes, such as IAXO, start to lose sensitivity near $m_a \gtrsim 0.01$eV. Further, the KSVZ model can now be probed at the direct detection experiments. A similar region of the $g_{a\gamma}-m_a$ space will also be investigated at LZ~\cite{Akerib:2019fml} and SuperCDMS SNOLAB~\cite{Agnese:2016cpb},
where the reach for $g_{a\gamma}$ needs to be scaled for the new detector type roughly by  $\sqrt{M_{D} Z^2_{D}/M_{Xe} Z^2_{Xe}}$ (where $M_D$ is the detector mass and $Z_D$ is the atomic number of the detector nucleus) for the same exposure.  \\

{\bf{\emph{Note Added:}}} During the completion of this work, a study~\cite{gao2020reexamining} appeared that also investigated the effect of the inverse Primakoff effect on solar axion detection. Also, another work~\cite{Abe:2012ut} appeared which discussed the impact that the choice of atomic form factor has on the inverse Primakoff cross section; we have since expanded our results to account for this variability.\\

\noindent {\bf \emph{Acknowledgements-}} The authors thank Nicole Bell, Kiwoon Choi, and Sebastian Hoof for discussions related to this work. JLN is supported in part by the Australian Research Council. JBD acknowledges support from the National Science Foundation under Grant No. NSF PHY182080. BD and AT acknowledge support from DOE Grant DE-SC0010813. AT also thanks the Mitchell Institute for Fundamental Physics and Astronomy for support. 

\appendix
\section{Solar Axion Flux}\label{sec:flux}
\begin{figure}
    \centering
    \includegraphics[width=0.45\textwidth]{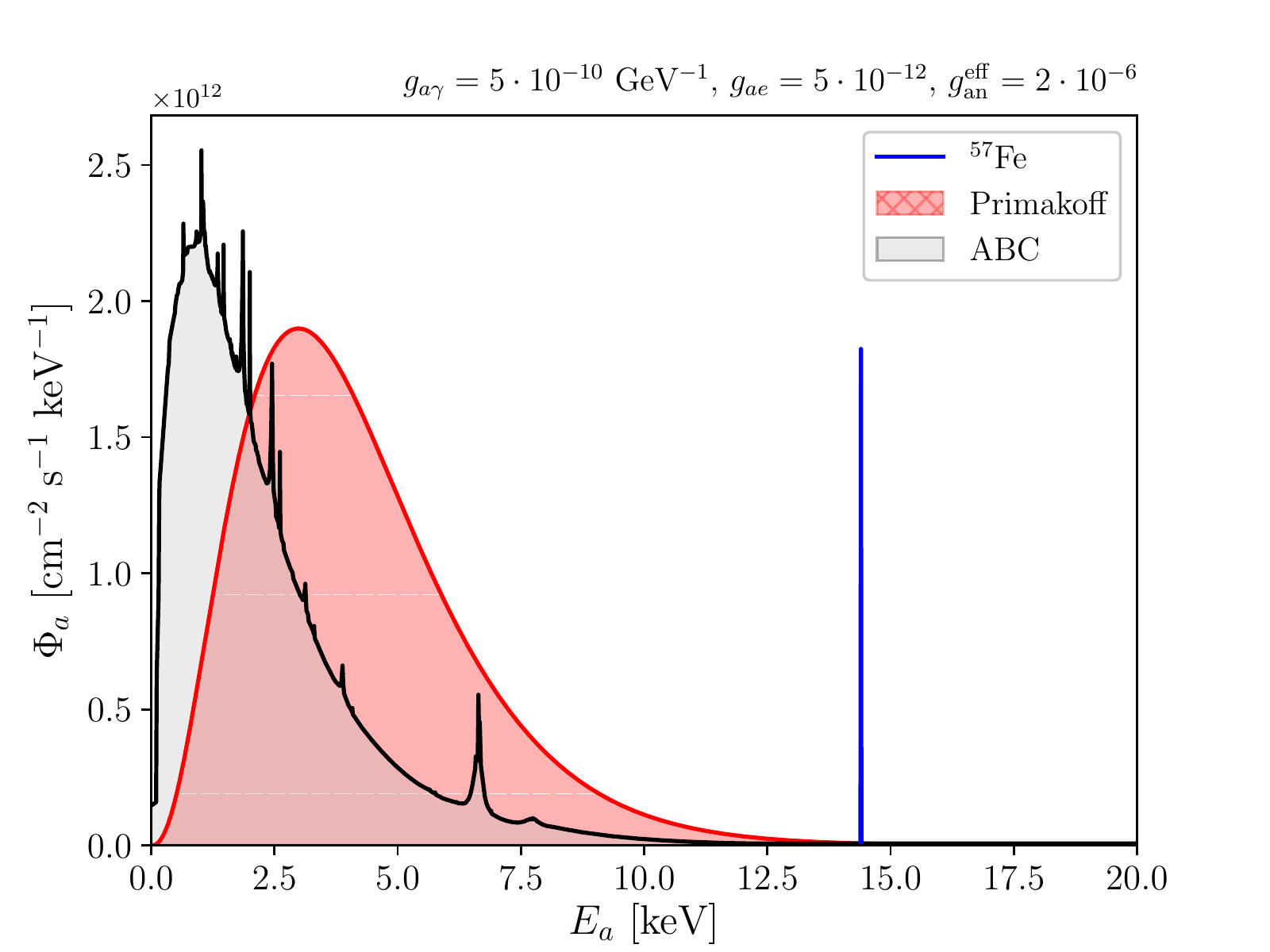}
    \caption{Solar axion fluxes at the Earth's surface are shown for the ABC, Primakoff, and $^{57}$Fe components. The bulk shape of the ABC component is due to axion-bremsstrahlung and Compton scattering, $e^- I \to e^- I\, a$ and $e^- \gamma \to e^- a$, while the numerous peaks are due to atomic transitions in $I^* \to I a$ and $e^- I \to I^- a$. The Primakoff flux exhibits a smooth thermal distribution from the coherent conversion of photons into axions, while the $^{57}$Fe flux is monoenergetic and is expected to broaden out from detector energy response effects.}
    \label{fig:fluxes}
\end{figure}

There are three sources of solar axion flux that we consider, each with a dependence on a different axion coupling parameter. First is the the ``ABC" flux, driving axion production from \textbf{A}tomic de-excitation and recombination, \textbf{B}remmstrahlung, and \textbf{C}ompton scattering processes~\cite{Redondo:2013wwa}, dependent on the $g_{ae}$ coupling. Next, the Primakoff production process, $\gamma Z \to a Z$, occurs via the $g_{a\gamma}$ coupling through the $t$-channel exchange of a virtual photon scattering with electrons or ions in the solar interior~\cite{Andriamonje:2007ew,Hagmann:2008zz}; this is essentially the process in Fig.~\ref{fig:scattering} under time reversal, modulo spin factors. Finally, de-excitation of $^{57}$Fe in the sun can produce a monoenergetic axion population at 14.4~keV~\cite{Moriyama:1995bz,Andriamonje:2009dx,Alessandria:2012mt}. This flux would arise from an effective axion-nucleon coupling $g_{\rm an}^{\rm eff} = -1.19g_{\rm an}^0 + g_{\rm an}^3$, where $g_{\rm an}^{0(3)}$ are the isoscalar (isovector) coupling constants for the nucleons \cite{Haxton:1991pu,Alessandria:2012mt}. Each of these flux components are shown in Fig.~\ref{fig:fluxes}.

\bibliography{main}

\end{document}